\documentclass{article}
\usepackage{amsmath, amssymb}
\usepackage{amsfonts}
\usepackage[mathscr]{eucal}
\usepackage{mathrsfs}

\def\fracd#1#2{{\frac{#1}{#2}}}

\def\eps{\varepsilon}

\def\d{\frac{d}{d t}}

\def\hf{\frac12}

\def\Om{{\Omega}}
\def\om{{\omega}}
\def\si{{\sigma}}
\def\a{\alpha}
\def\eps{\varepsilon}

\def\t{\tau}

\def\ga{\gamma}

\def\frA{\mathfrak{A}}
\def\cA{\mathcal{A}}
\def\cB{\mathcal{B}}
\def\sB{\mathscr{B}}

\def\sG{\mathscr{G}}

\def\sD{\mathscr{D}}

\def\sL{\mathscr{L}}
\def\cE{\mathcal{E}}
\def\H20{\mathcal{H}_{2,0}}

\def\cH{\mathcal{H}}
\def\cW{\mathcal{W}}

\def\R{\mathbb{R}}

\newcommand{\coH}{{\overline{H}}}
\newcommand{\wn}{{\widetilde{n}}}
\newcommand{\wrt}{{with respect to }}
\newcommand{\g}{{\nabla}}
\newcommand{\wE}{{\widetilde{E}}}
\newcommand{\wG}{{\widetilde{G}}}
\newcommand{\hn}{{\widehat{n}}}
\newcommand{\hE}{{\widehat{E}}}

\newtheorem{theorem}{Theorem}[section]
\newtheorem{lemma}[theorem]{Lemma}
\newtheorem{proposition}[theorem]{Proposition}
\newtheorem{definition}[theorem]{Definition}
\newtheorem{remark}[theorem]{Remark}

 \numberwithin{equation}{section}

\newenvironment{declaration}[1]{\trivlist
\item[\hskip \labelsep{\bf #1 }]\ignorespaces}{\endtrivlist}
\newenvironment{proofof}[1]{\begin{declaration}{#1}}{\hfill
$\square$ \end{declaration}}
\newenvironment{proof}{\begin{proofof}{Proof.}}{\end{proofof}}

\begin{document}

\title
 {Quantum  Zakharov Model in a Bounded Domain}

\author{Igor Chueshov\footnote{\small e-mail:
 chueshov@univer.kharkov.ua}
\\
Department of Mechanics and Mathematics,   \\  Kharkov
National University, \\ Kharkov, 61077,  Ukraine
}

\date{}

\maketitle

\begin{abstract}
We consider an initial boundary value problem for  a
quantum version of the  Zakharov system
arising in plasma physics. We prove the global well-posedness
of this problem in some Sobolev type classes and study properties of
 solutions.  This result confirms the conclusion
recently made in physical literature concerning
the absence of collapse in the quantum Langmuir waves.
In the dissipative case the existence of a
 finite dimensional global attractor
is  established and regularity properties
of this attractor are studied. For this we use the recently
developed method of  quasi-stability estimates.
In the case when external loads are
$C^\infty$ functions we show that every trajectory  from the attractor
is $C^\infty$ both in time and spatial variables.
This can be interpret as the absence of sharp coherent
structures in the limiting dynamics.
\end{abstract}

\section*{Introduction}
In a bounded domain $\Om \subset \R^d$, $d\le 3$,
we consider the following  system
\begin{equation}\label{QZ-1}
\left\{\begin{array}{l}
n_{tt}-\Delta\left(n+|E|^2\right)+h^2\Delta^2 n+\alpha n_t  =f(x),
\quad x\in\Om,\; t>0,
\\ \\
i E_t+\Delta E- h^2\Delta^2E  +i\gamma E - n E=g(x),\quad x\in\Om,\; t>0.
\end{array}\right.
\end{equation}
Here $E(x,t)$ is a complex function
and $n(x,t)$ is a real one, $h>0$,  $\alpha\ge 0$
 and $\gamma\ge 0$ are  parameters and $f(x)$, $g(x)$ are given
 (real and complex) functions.
\par
This system in dimension  $d=1$ was
obtained  in \cite{GHGO-2005}, by use of a quantum fluid approach,
to model the nonlinear interaction between quantum
Langmuir waves and quantum ion-acoustic waves in an
electron-ion dense quantum plasma.
Later a vector 3D version of equations (\ref{QZ-1}) was suggested  in \cite{HaShu-2009}.
In dimension $d=2,3$ the system in (\ref{QZ-1}) is
also known (see, e.g., \cite{SSS-2009} and the references therein)
as a simplified "scalar model"
which is in a good agreement with the vector model derived in
\cite{HaShu-2009}
 (see a discussion in \cite{SSS-2009}).
The quantum parameter $h$  in (\ref{QZ-1})
is proportional to the  Plank constant and
 expresses the ratio between the ion plasmon energy
and the electron thermal energy.
In the case when $h=0$ we arrive  to the classical (non-quantum)
Zakharov system which was introduced in
  \cite{Zakharov}  (also with $\a=\ga=0$,
 $f(x)\equiv 0$, $g(x)\equiv 0$)
 for the description of
wave phenomena in plasma.
This classical model was studied by many authors.
For well-posedness issues in the non-dissipative case  ( $\a=\ga=0$)
for $\Om\equiv\R^d$ with $d\le 3$ we refer
\cite{Bourgain1,Bourgain2,Ginibre,Glangetas}
and to the literature cited there.
In a bounded domain with $d=1,2$  well-posdeness
and long-time dynamics
were studied in \cite{ChuSch05,Flahaut,Goubet,Shcherbina}.
\par
Our main goal in this paper is to study well-posedness and long-time
issues  for the model in (\ref{QZ-1}).
Presence of the biharmonic operators in (\ref{QZ-1}) provides
additional a priori estimates and
makes it possible to study the model in all dimensions $d\le 3$
in a unified way.
For the sake of some simplification
 we consider   problem (\ref{QZ-1})
with the following  boundary conditions
\begin{equation}\label{Z-1bc-d}
n(x,t)=\Delta n(x,t)=0,~~ E(x,t)= \Delta E(x,t)=0~~\mbox{ for}~~ x\in\partial\Om,\; t>0.
\end{equation}
We can also consider other types of boundary
conditions related with the biharmonic operator
such as Dirichlet,  Neumann or periodic (if the domain is
rectangle) or else their combination.
Moreover, instead of the scalar amplitude $E(x,t)$
we can take a complex vector $(E_1(x,t);\ldots; E_d(x,t))$ satisfying
the corresponding (vector) Schr\"odinger equation. This model describes
the so-called electrostatic limit of the original vector model
(see the discussion in \cite{SSS-2009}). To avoid extra technicalities
we do not pursue these generalizations in this paper.
\par
In this paper we prove  (see Theorem~\ref{th:wp})
that the Cauchy problem for (\ref{QZ-1}) equipped
with  the boundary conditions in (\ref{Z-1bc-d})  has
a unique weak solution
for initial data $(n_1; n_0; E_0)$ from the natural  phase space.
To prove the existence of the solutions we use Galerkin approximations
and the standard compactness method.
We also show  that these solutions satisfy some energy type relations
and study their regularity properties.
In particular we
 prove the Lipschitz continuous dependence of solutions
on initial data and
 show that these solutions generate a continuous semiflow $S_t$.
 This global well-posedness  result confirms the
physically motivated  conclusion
recently made in \cite{HaShu-2009} and \cite{SSS-2009}
 concerning
the absence of collapse in the quantum Langmuir waves.
\par
In the dissipative case ($\alpha>0$, $\gamma>0$) we study long-time
behaviour of the semiflow $S_t$ generated by weak solutions
 and prove the existence of a compact global attractor.
This attractor has finite fractal dimension and
attracts smooth (semi-strong) solutions in a stronger topology.
We also show that all trajectories from the attractor
are $C^\infty$ in time variable. Moreover, if the external forces
$f$ and $g$ are $C^\infty$ functions,
then  elements from the attractor are also $C^\infty$
in the spatial variables.
This can be interpret as the absence of sharp coherent
structures in the long-time limit.
To prove these results (see Thorem~\ref{t:attr})
we use a combination of the traditional approach
(see, e.g., \cite{BV92,Chu99,Lad91,Temam}) with
the recently developed method
\cite{ChuLas_JDDE_2004,ChuLas,cl-book} based
on quasi-stability properties of the system.
To study  regularity of solutions on the attractor we also
use some ideas from \cite{GT87}.
\par
The paper is organized as follows.
In Section~\ref{sect2} we state and discuss our main results
on well-posedness (Theorem~\ref{th:wp}) and on long lime dynamics
(Theorem~\ref{t:attr}).
Sections~\ref{sect3} and \ref{sec:glob-at}
are devoted to the  proof of Theorem~\ref{th:wp} and
  Theorem~\ref{t:attr} respectively.

\section{Statements of the main results}\label{sect2}
Let $\Om\subset\R^d$ be  a smooth
bounded domain, $d\le 3$,  and
$H^s(\Om)$ be $L_2$ based Sobolev space, $s\in\R$.
We also denote by $H_0^s(\Om)$ the completion of
$C_0^\infty(\Om)$ in  $H^s(\Om)$.
Let
 $A$ be the
minus Dirichlet Laplace operator:
\[
Au=-\Delta u ~~~\mbox{for any}~~
u\in \sD(A)=(H^2\cap H^1_0)(\Om).
\]
The operator $A$ is a self-adjoint positive
 operator in $H=L_2(\Om)$  and has
a compact inverse. The latter implies
the existence of  an orthonormal basis
 $\{e_k\}$  in $H$ consisting of eigenfunctions of $A$:
\[
Ae_k=\lambda_k e_k,\quad 0<\lambda_1\le\lambda_2\le\ldots,\quad
\lim_{k\to\infty}\lambda_k=\infty.
\]
We define the Sobolev type spaces $H_s$ by the formula
$H_s=\sD(A^{s/2})$ with the graph norm $\|\cdot\|_s\equiv\|A^{s/2}\cdot\|$
for $s\ge 0$. If $s<0$, then  $H_{s}=\left[H_{-s}\right]'$ is the completion
of $H_0\equiv H=L_2(\Om)$ \wrt $\|\cdot\|_{s}\equiv\|A^{s/2}\cdot\|$.
From interpolation (see, e.g., \cite{Lions}) we have  that
(a)~$H_s\subset H^s(\Om)$ for $s\ge 0$,
(b)~$H_s=(H^s\cap H^1_0)(\Om)$ for $s\in[1,2]$, $s\neq 3/2$,
and (c)~$H_s=(H^s_0)(\Om)$ for $s\in [0,1]$, $s\neq 1/2$.
\par
We  mention that we deal with both real and complex versions
of the spaces $H_s$. We keep notation $H_s$ for real case
and denote by $\coH_s$ its complexification. For norms and inner products
we use the same notations.
\par
We recall (see, e.g., \cite{Triebel78}) for further use the following
Sobolev embeddings:
\begin{equation}\label{sobolev}
H^{s}(\Om)\subset L_p(\Om) ~\mbox{with}~
s=d\left(\frac12 -\frac1p\right),~ p\ge 2,
~~\mbox{and}~~ H^{s}(\Om)\subset C(\Om)
 ~\mbox{for} ~ s>\frac{d}2.
\end{equation}
In particular, since  $d\le3$, both spaces $H_2$ and $\coH_2$
are Banach algebras.

We also introduce the (phase) space
$\cH= H\times H_2 \times \coH_2$
endowed with the norm
\begin{equation}\label{cH-norm}
\|(n_1;n_0;E_0\|_\cH^2=  \|n_1\|^2+\|n_0\|_2^2+\|E_0\|_2^2.
\end{equation}
 Now we rewrite  problem (\ref{QZ-1}) with the boundary conditions in
(\ref{Z-1bc-d}) in  the following   form:
\begin{equation}
\label{Z_nd1}
    n_{tt}+A \left(n+ |E|^2\right) +h^2 A^2 n+\alpha n_t  =f,
\end{equation}
\begin{equation}
\label{Z_nd2}
    i E_t-A E - h^2 A^2 E- n E+i\gamma E =g.
\end{equation}
We equip equations (\ref{Z_nd1}) and (\ref{Z_nd2})
with initial data
\begin{equation}
\label{Z_nd3}
    n_t|_{t=0}=n_0,\; n|_{t=0}=n_1,\; E|_{t=0}=E_0.
\end{equation}
Below we  assume that (i) $h>0$, (ii)  $\alpha,\gamma\ge 0$,
(iii) the real
$f(x)$ and the complex $g(x)$ functions are given
from  $L_2(\Om)$.
\par
We understand solutions to problem
(\ref{Z_nd1})--(\ref{Z_nd3}) in the sense of the following
definition.
\begin{definition}\label{de:weak-sol}
{\rm
A pair  $(n;E)$ is said to be a \emph{weak} solution to
problem
(\ref{Z_nd1})--(\ref{Z_nd3})
on an interval $[0,T]$ iff
\begin{equation*}
(n_t; n; E)\in  L_\infty\left([0,T];\,  H\times H_2 \times \coH_2
 \right),
\end{equation*} and
(i) relations (\ref{Z_nd1}) and (\ref{Z_nd2}) are satisfied in the sense
of distributions,
(ii) initial data (\ref{Z_nd3}) hold.
}
\end{definition}
Since by (\ref{sobolev})  $nE\in \coH_2$ and $|E|^2\in H_2$ for
$E\in \coH_2$ and $n\in H_2$,
it follows from
 (\ref{Z_nd1}) and (\ref{Z_nd2})
 that
\begin{equation}\label{Z-def1}
n_{tt}\in  L_\infty\left([0,T];\,  H_{-2}\right)
\quad\mbox{and}\quad E_{t}\in
L_\infty\left([0,T];\, \coH_{-2}
\right)
\end{equation}
 for any weak solution $(n;E)$.
In particular, this means that any weak solution
$(n;E)$ satisfies  (\ref{Z_nd1}) and (\ref{Z_nd2})  for almost all
$t\in [0,T]$ as equalities in $H_{-2}$ and $\coH_{-2}$.
Moreover,  by Lions' lemma (see~\cite[Lemma~8.1]{Lions})
the triple $(n_t; n; E)$ is a weakly continuous function with values
in $H\times H_2 \times \coH_2$ and hence
\begin{equation}\label{Z-def2}
(n_t; n; E_t; E)\in  C\left([0,T];\, H_{-\si}\times H_{2-\si}
  \times \coH_{-2-\si}\times \coH_{2-\si}\right),~~~\forall\,\si>0.
\end{equation}
\par
 Our first result is the following well-posedness theorem
for weak solutions to problem (\ref{Z_nd1})--(\ref{Z_nd3}).
\begin{theorem}\label{th:wp}
Let $\alpha,\ga\ge 0$ and $h>0$.
Assume that
$(n_1;n_0;E_0)\in\cH$,  $f\in H$ and $g\in\coH$.
Then problem
(\ref{Z_nd1})--(\ref{Z_nd3}) on every interval $[0,T]$
has a unique weak solution.
This solution possesses the property
\begin{equation}\label{Z-def2-0}
(n_t; n; E_t; E)\in  C\left([0,T];\, H\times H_{2}
  \times \coH_{-2}\times \coH_{2}\right)
\end{equation}
and satisfies  the  relations
\begin{equation}
\label{ex_diss_3a}
\|E(t)\|\le \|E_0\| e^{-\gamma t}+
e_\ga(t)\cdot \|g\|~~~\mbox{for any}~~ t\ge 0,
\end{equation}
where $e_\ga(t) =\gamma^{-1} \left(1-e^{-\gamma t}\right)$
in the case $\gamma>0$ and $e_\ga(t)=t$
if $\gamma= 0$. \par
We also have that
\begin{enumerate}
    \item[{\bf (1)}] There exists a constant $C_{T,R}>0$ such that
for any couple of initial data
 $Y^i=(n^i_1;n^i_0;E^i_0)\in\cH$, $i=1,2$, with the property
 $\|Y^i\|_\cH\le R$  we have the relation
\begin{equation}\label{lip}
\|Y^1(t)-Y^2(t)\|_\cH\le C_{T,R} \|Y^1-Y^2\|_\cH, ~~~ t\in [0,T],
\end{equation}
where $Y^i(t)=(n^i_t(t);n^i(t);E^i(t))$ is the solution which corresponds
to the initial data  $Y^i=(n^i_1;n^i_0;E^i_0)$.
The norm $\|\cdot\|_\cH$ is given by (\ref{cH-norm}).
 \item[{\bf (2)}]  For any $t\in [0,T]$ we have
the following energy balance relation
\begin{multline}\label{w-en2}
V_0(n_t(t),n(t)) +V_1(n(t),E(t))
 +\int_0^t\left(\alpha\|n_t\|_{-1}^2+
2\gamma V_1(n,E)\right)d\tau
  \\
  =
V_0(n_1,n_0)+V_1(n_0,E_0)
+2\gamma\int_0^t  \Re(g, E) d\tau,
\end{multline}
where
 $V_0(n_t,n)$ is given by
\begin{equation}\label{Zen-n1-2}
V_0(n_t,n)    =\frac12\left[
\|n_t\|^2_{-1}+\| n\|^2 +h^2 \| n\|_1^2\right] -(f,n)_{-1}
\end{equation}
and
\begin{equation}\label{r-en-2}
V_1(n, E)=\|E\|_1^2 +h^2 \|E\|_2^2  +(n, |E|^2)+2\Re (g,E).
\end{equation}
\item[{\bf (3)}]  If in addition we assume that  $E_0\in H_{4}$, then
the corresponding weak solution is   semi-strong, i.e.,
\begin{equation}\label{Z-def2-0sm2}
(n_t; n; E_t; E)\in  C\left([0,T];\, H\times H_{2}
  \times \coH\times \coH_{4}\right).
\end{equation}
This solution is Lipschitz on $\cH_*=H\times H_2\times \coH_4$
with respect to initial data, i.e.,
there exists a constant $C_{T,R}>0$ such that
\begin{equation}\label{lip-h*}
\|Y^1(t)-Y^2(t)\|_{\cH_*}\le C_{T,R} \|Y^1-Y^2\|_{\cH_*}, ~~~ t\in [0,T],
\end{equation}
where $Y^i(t)=(n^i_t(t);n^i(t);E^i(t))$ is the solution which corresponds
to the initial data  $Y^i=(n^i_1;n^i_0;E^i_0)\in\cH_*$
such that $\|Y^i\|_{\cH_*}\le R$.
Moreover,
the following energy type relation
\begin{align}\label{w-en}
\cE_f(n_t(t),n(t)) &+\|E_t(t)\|^2 +\int_0^t\left(\alpha\|n_t(\tau)\|^2+
2\gamma \|E_t(\tau)\|^2\right)d\tau
 \nonumber \\
 &  =
\cE_f(n_1,n_0)+\|E_1\|^2+2\int_0^t R(n_t(\tau),n(\tau), E(\tau))d\tau
\end{align}
holds
for any $t\in [0,T]$,
where $\cE_f(n_t,n)$ is given by
\begin{equation}\label{Zen-n1}
\cE_f(n_t,n)    =\frac12\left[ \|n_t\|^2+\|A^{1/2} n\|^2 +h^2\|A n\|^2
-2(f,n)\right],
\end{equation}
the element $E_1\in\coH$ is defined by $n_0$ and $E_0$ from (\ref{Z_nd2}), i.e.
\begin{equation}\label{Zen-E1}
E_1=   - i \left( A E_0+ h^2A^2 E_0 + n_0 E_0-i\gamma E_0 +g\right),
\end{equation}
and
$R(n_t,n, E)=(n_t, |\g E|^2)+\Re (n_t, iE_t +\Delta E)$.
\end{enumerate}
\end{theorem}
In particular Theorem~\ref{th:wp} allows us to define
an evolution (semigroup) operator $S_t$ by the formula
\begin{equation*}
S_tY=Y(t)\equiv (n_t(t);n(t);E(t)),~~~ Y=(n_1;n_0;E_0),
\end{equation*}
where $(n(t);E(t))$ is a weak solution to  problem
(\ref{Z_nd1})--(\ref{Z_nd3})
with  the initial data  $Y=(n_1;n_0;E_0)$.
This evolution operator generates dynamical systems
$(\cH,S_t)$ and  $(\cH_*,S_t)$ with the phase spaces
$\cH$ and $\cH_*$.
Long-time dynamics of these systems
is the main point of interest in this paper.
\par
We  recall (see, e.g., \cite{BV92,Chu99,Temam})
that the \textit{global attractor}  of a dynamical system
$\big( X, S_t\big)$ in a Banach space $X$
is defined as a bounded closed  set $\frA\subset X$
which is  invariant ($S(t)\frA=\frA$ for  $t>0$)
and  uniformly  attracts
all other bounded  sets:
$$
\lim_{t\to+\infty} \sup\{{\rm dist}_X(S_ty,\frA):\ y\in B\} = 0
\quad\mbox{for any bounded  set $B$ in $X$.}
$$
\par
Our main results   on asymptotic dynamics
of the dynamical system generated by (\ref{Z_nd1})--(\ref{Z_nd3})
are collected in the following assertion.

\begin{theorem}\label{t:attr}
 Assume that $\alpha,\ga,h >0$ and $(f;g) \in H\times \coH$.
Then
the dynamical
system $(\cH,S_t)$ generated by problem   (\ref{Z_nd1}) -- (\ref{Z_nd3})
has a compact global attractor $\frA$.
This attractor possesses the following properties.
\begin{enumerate}
    \item[{\bf (1)}] The   fractal dimension ${dim}_f\frA$ of  $\frA$
is finite.
  \item[{\bf (2)}]  $\frA$ is a bounded set in
$H_2\times H_4\times\coH_4$.
  \item[{\bf (3)}]  $\frA$ is also  a global attractor for
the system $(\cH_*,S_t)$, i.e., it uniformly attracts
bounded sets from  $\cH_*\equiv H\times H_2\times\coH_4$ in the topology of $\cH_*$.
 \item[{\bf (4)}]
For any trajectory
$U(t)=(n_t(t);n(t);E(t))$,  $-\infty<t<+\infty$,
from the attractor we have that
$(n(t);E(t))\in C^\infty(\R; H_4\times\coH_4)$ and
\begin{equation}\label{U-atr-bnd}
\sup_{t\in\R}\left\{\|n^{(m)}(t)\|^2_4+\|E^{(m)}(t)\|^2_4  \right\}
\le R_m, \quad \forall\, m=0,1,2,\ldots,
\end{equation}
where $n^{(m)}(t)$ and $E^{(m)}(t)$ denote time derivatives
of the order $m$.
  \item[{\bf (5)}] If $f,g$ are $C^\infty$ functions on $\overline{\Om}$,
then  $\frA\subset C^\infty(\overline{\Om})\times  C^\infty(\overline{\Om})\times\overline{C}^\infty(\overline{\Om})$
and
\[
\sup_{t\in\R}\max_{x\in\overline{\Om}}
\left\{|\nabla^l_x n^{(m)}(x,t)|+|\nabla^l_x E^{(m)}(x,t)|  \right\}
\le R_{ml}, \quad \forall\, m,l=0,1,2,\ldots,
\]
for any trajectory
$U=(n_t;n;E)$ from the attractor.
\end{enumerate}
\end{theorem}
We prove this theorem in Section~\ref{sec:glob-at}.

\section{Proof of Theorem~\ref{th:wp}}\label{sect3}

We use  the compactness method and split the proof in several steps.

\subsection{Related linear problem}

 We first present  two preliminary
assertions  concerning related linear problems.
\begin{lemma}\label{le:n-lin}
Let  $(n_1;n_0)\in H\times H_2$ and   $F(t)\in L_2(0,T; H)$.
Then there exists a unique weak solution  $n(t)$ to the linear problem
\begin{equation}
\label{n-lin}
    n_{tt}+(A  +h^2 A^2) n+\alpha n_t  =F(t),
 ~~t>0,~~~n(0)=n_0,~ n_t(0)=n_1.
\end{equation}
This solution possesses the property
 $(n_t;n)\in C([0,T]; H\times H_2)$ and  satisfies the following
energy relation
\begin{equation}\label{lin-en-n0}
\cE_0(n_t(t),n(t))+\a \int_0^t\|n_t(\tau)\|^2d\tau
 =
\cE_0(n_1,n_0)+\int_0^t (F(\t),n_t(\tau))d\tau
\end{equation}
 for every $t\in [0,T]$, where the energy $\cE_0(n_t,n)$
is given by (\ref{Zen-n1}) with $f=0$, i.e.,
\begin{equation}\label{lin-en-n1}
\cE_0(n_t,n)    =\frac12\left[ \|n_t\|^2+\|A^{1/2} n\|^2 +h^2\|A n\|^2
\right].
\end{equation}
Moreover, in the case $F(t)\equiv 0$ equation (\ref{n-lin})
defines a $C_0$-semigroup $T_t$  in  each space $H_\si\times H_{2+\si}$,
$\si\in \R$,
by the formula
\[
T_tw =  (n_t(t);n(t)),~~~ w=(n_1;n_0),
\]
where $n(t)$ solves  (\ref{n-lin}) with $F\equiv 0$. This semigroup is exponentially
stable in $H_\si\times H_{2+\si}$, i.e.,
there exist $M,\om>0$ such that
\begin{equation}\label{T-exp-st}
\|T_tw\|_{H_\si\times H_{2+\si}}\le
Me^{-\om t}\|w\|_{H_\si\times H_{2+\si}},
~~~t\ge 0,~~\forall\, w\in H_\si\times H_{2+\si}.
\end{equation}
\end{lemma}
 \begin{proof}
The first part of the statement
 follows from the results given in \cite[Chap.3, Theorem~8.2]{Lions}.
As for the second part it follows from the standard calculations
involving the Lyapunov function
\[
\sL(n_t;n)=\hf\left[ \|n_t\|_\si^2 + \|n\|_{1+\si}^2+h^2\|n\|^2_{2+\si}
\right]
+\eps\left[ ( A^{\si}n,n_t)+
\frac{\alpha}2 \|n\|_\si^2\right]
\]
with appropriate $\eps>0$, see, e.g., \cite[Chapter 4]{Temam}.
\end{proof}
\begin{lemma}\label{le:schr-lin}
Let $E_0\in\coH_2$, $G(t)\in C([0,T]; \coH)$ with $G_t\in L_2(0,T; \coH_{-2})$.
Assume that $a(t)\in L_2(0,T; H_2)\cap C^1(0,T; H_{-2})$.
Then  the following Cauchy  problem for the
non-autonomous Scr\"{o}dinger equation
\begin{equation}
\label{schr-lin1}
    i E_t-(A + h^2 A^2) E -a(t)E +i\gamma E =G(t),~~t>0,~~~
E(0)=E_0,
\end{equation}
has a unique weak solution $E(t)$ which belongs to
$C(0,T; \coH_2)\cap C^1(0,T; \coH_{-2}) $
and satisfies the following balance relations:
\begin{equation}\label{en-E-lin1}
\|E(t)\|^2+2\gamma \int_0^t\|E(\tau)\|^2d\tau
=
\|E_0\|^2
+2\Im \int_0^t (G(\tau), E(\tau))d\tau
\end{equation}
and
\begin{equation}\label{en-Elin2}
W(t)+2\gamma \! \int_0^t \!\! W(\t) d\tau
\\
=
W(0)
+\Re\! \int_0^t (a_t(\tau) E(\tau) +2G_t(\tau)+ 2\gamma G(\t), E(\tau))d\tau,
\end{equation}
where
\begin{equation*}
W(t)= \|E(t)\|_1^2+h^2\|E(t)\|_2^2 + (a(t),|E(t)|^2)+2 \Re (G(t),E(t)).
\end{equation*}
\par
If in addition $E_0\in \coH_4$, $a_t\in  L_2(0,T; H)$ and
$G_t\in  L_2(0,T; \coH)$, then
$E(t)$  belongs to the space
$C(0,T; \coH_4)\cap C^1(0,T; \coH) $ and the following balance relation
\begin{equation}\label{Zen-E0lin}
\|E_t(t)\|^2 +2\ga  \int_0^t \|E_t(\tau)\|^2d\tau =
\|E_1\|^2
+2\Im \int_0^t (a_t(\tau) E(\tau)+G_t(\tau), E_t(\tau))d\tau
\end{equation}
holds with
 $E_1=   - i \left( (A +h^2 A^2) E_0 + a(0) E_0-i\gamma E_0 +G(0)\right)$.
\end{lemma}
\begin{proof}
We apply the results from
 \cite[Chap.3, Sect.10]{Lions}.
For this we consider the following sesqui\-linear form
\[
a(t,u,v)= (A u,v)+h^2(Au,Av)+\int_\Om a(t,x) u(x)\overline{v(x)}\, dx
\]
on $\coH_2$.  The form $a(t,u,v)$ is bounded and
 coercive  in the  sense that there exist positive $\om_1$
and $\om_2$ such that
\[
a(t,u,u)+\om_1\|u\|^2\ge \om_2\|u\|_2^2,\quad u\in \coH_2.
\]
 It is also continuously differentiable \wrt $t$. Therefore,
 by \cite[Theorem~10.1]{Lions} problem (\ref{Z_nd2wd}) is
has a unique weak solution which is weakly continuous in $\coH_2$.
The relations in (\ref{en-E-lin1}) and (\ref{en-Elin2})
follows by the standard argument
with the help of the multipliers $P_NE$ and $P_NE_t$
and the subsequent limit transition $N\to\infty$
(we refer to \cite{Lions} and to similar calculations
below in the case of the nonlinear model in (\ref{QZ-1})).
 The strong continuity of $E(t)$
follows from weak continuity and from (\ref{en-Elin2}),
see also Remark~10.2 in \cite{Lions}.
\par
To prove the second part of Lemma~\ref{le:schr-lin} we note that
the function $\widetilde{E}\equiv E_t$ is a solution (in the sense
of distributions) to the problem
\begin{equation}
\label{Z_nd2wd}
    i \wE_t -(A  +h^2 A^2)\wE- a(t) \wE+ i\gamma \wE
=\widetilde{G}(t)\equiv a_t E+G_t,
    \quad \wE|_{t=0}=E_1,
\end{equation}
where $E_1=   - i \left( A E_0+h^2 A^2 E_0
+ a(0) E_0-i\gamma E_0 +G(0)\right)$.
Since $E_1\in\coH$ and $\wG(t)\in L_2(0,T;\coH)$,
 by \cite[Theorem~11.1]{Lions} problem (\ref{Z_nd2wd}) is uniquely solvable
 in the sense of distributions.
\par
Let $\{\wE^N_1\}\subset\coH_2$ and $\{\widetilde{G}^N(t)\}\subset C^1(0,T;\coH)$
be sequences such that
\[
\lim_{N\to\infty}\left\{\|\wE^N_1-E_1\|+ \int_0^T\| \wG^N(t)-\wG(t)\|^2 dt
\right\}=0.
\]
Consider the problem
\[
 i \wE^N_t- (A+h^2A^2) \wE^N - a(t) \wE^N =G^N(t),    \quad \wE|_{t=0}=\wE^N_1.
\]
By the first part of the lemma this problem has a unique solution
\[
\wE^N(t)\in  C(0,T;\coH_2)\cap C^1(0,T;\coH_{-2})
\]
which satisfies the relations
\begin{equation}\label{en-E-N}
\|\wE^N(t)\|^2 +2\ga \int_0^t \|\wE^N(\tau)\|^2d\tau =
\|\wE^N_1\|^2
+2\Im \int_0^t (G^N(\tau),\wE^N(\tau))d\tau
\end{equation}
and
\begin{equation}\label{en-E-N1}
\|\wE^{N_1}(t)-\wE^{N_2}(t)\|^2\le
\|\wE^{N_1}_1-\wE^{N_2}_1\|^2
+2\Im \int_0^t (G^{N_1}-G^{N_2},\wE^{N_1}-\wE^{N_2})d\tau
\end{equation}
for any $N$, $N_1$ and $N_2$. By Gronwall's lemma from (\ref{en-E-N1})
we have that
\[
\max_{[0,T]}\|\wE^{N_1}(t)-\wE^{N_2}(t)\|^2\to0
\quad\mbox{as}\quad N_1,N_2\to\infty.
\]
Thus there exists a function $E^*(t)\in C(0,T;\coH)$ such that
\[
\lim_{N\to\infty}\max_{[0,T]}\|\wE^{N}(t)-E^*(t)\|^2=0.
\]
By the uniqueness of solutions to  problem (\ref{Z_nd2wd})
with a fixed $\wG(t)$
we can conclude that $E^*(t)\equiv \wE(t)$. Thus
$E_t(t)\equiv \wE(t)\in  C(0,T;\coH)$. Consequently
using equation (\ref{schr-lin1})
 we obtain the required regularity of $E$.
\par
Relation (\ref{Zen-E0lin}) follows
after the limit transition in (\ref{en-E-N}).
\end{proof}
\subsection{Existence}\label{sect4}
 We  start with the Galerkin  approximations  of the Zakharov problem:
\begin{equation}
\label{exist_1a}
\left\{\begin{array}{l}
n_{tt}^N +A\left(n^N+P_N |E^N|^2\right)+h^2A^2 n^N+\alpha n^N_t=
 P_N f,\\ [2mm]
i E_t^N- A E^N -h^2 A^2 E^N-P_N(n^N E^N)+i\gamma E^N =P_N g.
\end{array}\right.
\end{equation}
Here $P_N$ is the orthoprojector
on Span$\{e_k\, :\, k=1,2,\ldots,N\}$, where $\{e_k\}$ is
the eigenbasis of $A$. The functions $n^N$ and $E^N$ have
their values in $P_NH$ and  $P_N\coH$.
To simplify notations we omit below the index $N$.
We use the same standard multipliers
as for the  classical Zakharov system
(see, e.g., \cite{ChuSch05} and the references therein).
\par
Multiplying in $\coH$ the second equation
from (\ref{exist_1a}) by $E$
and taking  the imaginary part
we obtain that
$$
\hf \d\|E\|^2+\gamma \|E\|^2=\Im(g, E)\le
\|E\| \|g\|.
$$
From this relation we conclude that
relation (\ref{ex_diss_3a})
 holds for this approximate solution.
 Thus we have that
\begin{equation}
\label{exist_2b}
\|E(t)\|\le K(E_0,g, T)\equiv K,\quad t\in [0,T].
\end{equation}
We  multiply the first equation from (\ref{exist_1a})
 by $A^{-1}n_t$ in $H$ and obtain that
\begin{equation}
\label{exist_3a}
 \d V_0(n_t,n)+(n_t,|E|^2)+ \alpha \|n_t\|^2_{-1}=0,
\end{equation}
where $V_0(n,n_t)$ is given by (\ref{Zen-n1-2}).
Now we multiply in $\coH$ the second equation from (\ref{exist_1a})
by $2E_t+2\gamma E$ and take the real part of the result.
 One can see that
\begin{equation}
\label{exist_4a}
  \d V_1(n,E)  + 2\gamma(
\|E\|_1^2 +h^2\|E\|_2^2)-(n_t,|E|^2)=\Psi(t),
\end{equation}
where  $V_1(n,E)$ is given by (\ref{r-en-2}) and
$\Psi(t)=
 -2\gamma\left[  (n,|E|^2)+\Re (g, E)\right]$.
Taking the sum of (\ref{exist_3a}) and (\ref{exist_4a}), we get
\begin{equation}
\label{exist_5a}
\d V(n(t),n_t(t),E(t))+2\gamma\left( \| E\|_1^2+h^2\|E\|_2^2\right)  + \alpha \|n_t\|^2_{-1}
=\Psi(t),
\end{equation}
where  $V(n,n_t,E)=V_0(n_t,n)+V_1(n,E)$, i.e.,
\begin{multline*}
V(n,n_t,E)= \hf\left[\|n_t\|^2_{-1}+\|n\|^2+h^2 \|n\|^2_1\right]
-(f,n)_{-1}          \\   +\|E\|_1^2 +h^2\|E\|_2^2
+2\Re(g, E)+(n,|E|^2).
\end{multline*}
Let
\begin{equation}
\label{exist_5b+}
V_+(n,n_t,E)= \hf\left[\|n_t\|^2_{-1}+\|n\|^2+h^2 \|n\|^2_1\right]
          +\| E\|_1^2 +h^2\|E\|_2^2.
\end{equation}
Then there exists $a_i,b_i>0$ such that
\begin{equation}
\label{v-v+}
a_0V_+(n,n_t,E)-a_1(1+ K^{10})\le
V(n,n_t,E)\le b_0V_+(n,n_t,E)+b_1(1+ K^{10})
\end{equation}
for all $\alpha,\ga\ge 0$, $h>0$, where $K$ is given in (\ref{exist_2b}).
Indeed, by the embedding in (\ref{sobolev}) with $p=4$
 we have that
\[
 |(n,|E|^2)| \le C \|n\| \|E\|^2_{L_4(\Om)}\le
 C \|n\| \|E\|^2_{3/4}.
\]
Thus by interpolation  using  (\ref{exist_2b}) we have that
\begin{align}
\label{exist_8a}
   |(n,|E|^2)| &\le C \|n\| \|E\|^{5/4} \|E\|^{3/4}_2
\le CK^{5/4}\|n\|  \|E\|^{3/4}_2
 \nonumber \\ &\le \eta( \|n\|^2+ \|E\|^2_2)+ C_\eta K^{10}
 \end{align}
for every $\eta>0$.   This
and  obvious estimates for the linear terms
imply (\ref{v-v+}).
\par
By the same argument we also have that
\[
\Psi(t)\le C_K\left(1+\|n(t)\|^2 +\|E(t)\|_2^2\right).
\]
Therefore
using Gronwall's lemma
 we obtain from  ~(\ref{exist_5a})
 the second a priori estimate:
\begin{equation}
\label{exist_9d}
\|n_t(t)\|^2_{-1}+\|n(t)\|_1^2+\|E(t)\|_2^2\le C_K(T, f,g),
\quad t\in [0,T].
 \end{equation}
Now we improve the estimates for $n$ and $n_t$.
Multiplying the first equation of~(\ref{exist_1a}) by~$ n_t$
yields
\begin{equation}
\label{exist_11a0}
\hf \d\left(\|n_t\|^2+\| n\|_1^2 +h^2\|n\|_2^2
-2(f,n)\right)
+(n_t,-\Delta |E|^2)+\alpha\|n_t\|^2=0,
\end{equation}
Since by    (\ref{exist_9d}) we have
$\|\Delta |E|^2\|\le C \|E\|_2^2\le C$ for all $t\in [0,T]$,
 via Gronwall's lemma we obtain from (\ref{exist_11a0}) that
\[
\|n_t\|^2+\|n\|_2^2\le C(T, f,g),
\quad t\in [0,T].
\]
Thus the approximate solutions $(n^N;E^N)$
possesses the following a  priori  estimate
\[
\|n^N_t(t)\|^2+\|n^N(t)\|_2^2+\|E^N(t)\|_2^2\le C(T, f,g),
\quad t\in [0,T].
 \]
This estimate allow us to make limit transition $N\to\infty$
and prove the existence of weak solutions
satisfying the relation
\begin{equation}
\label{exist_9d+1}
\|n_t(t)\|^2+\|n(t)\|_2^2+\|E(t)\|_2^2\le C(T,R),
\quad t\in [0,T],
 \end{equation}
for every initial data $Y_0=(n_1;n_0;E_0)\in \cH$ such that
$\|Y_0\|_\cH\le R$.

\subsection{Uniqueness and the Lipschitz property in $\cH$}

Now we prove the uniqueness and the Lipshcitz property in (\ref{lip}).
\par
Let $(n^i(t);E^i(t))$, $i=1,2$, be two weak solutions
with different initial data $Y^i=(n_1;n_0;E_0)$
from the ball $\cB_R=\{Y\in\cH : \|Y\|_\cH\le R\}$.
By Definition~\ref{de:weak-sol} we can assume that
\begin{equation}
\label{est-M}
\|n^i_t(t)\|^2+\|n^i(t)\|_2^2+\|E^i(t)\|_2^2\le M^2,~~~ t\in [0,T],
\end{equation}
for some $M>0$ (by (\ref{exist_9d+1}) for the
solution constructed in Section~\ref{sect4} the constant $M$ is
determined by $R$ and $T$).
 The   difference
\begin{equation}
\label{dif-def}
(n(t);E(t))=(n^1(t)-n^2(t);E^1(t)-E^2(t))
\end{equation}
solves  equations
\begin{equation}
\label{Z_nd1-df}
    n_{tt}+ \left(A+ h^2 A^2\right)n+\alpha n_t  =F(t),
\end{equation}
where $F(t)=\Delta( |E^1(t)|^2- |E^2(t)|^2)$, and
\begin{equation}
\label{Z_nd2-df}
    i E_t-(A + h^2 A^2) E -n^1(t)E +i\gamma E =G(t),
\end{equation}
where $G(t)=(n^1(t)-n^2(t))E^2(t)=n(t) E^2(t)$.
\par
It is clear the $F$ satisfies
the hypotheses of Lemma~\ref{le:n-lin} and
by (\ref{est-M})
\[
\|F(t)\|\le C_M\|E(t)\|_2,~~~ t\in [0,T].
\]
Thus Lemma~\ref{le:n-lin} yields
\begin{equation}\label{n-df}
\cE_0(n_t(t),n(t))
\le
\cE_0(n_t(0),n(0))+C_M\int_0^t
\left[ \|E(\t)\|_2^2+ \|n_t(\tau)\|^2\right]d\tau.
\end{equation}
By the Sobolev embeddings (\ref{sobolev}) from (\ref{Z-def2})
 we have that
 $G(t)\in C(0,T; \coH)$. Moreover,
from (\ref{Z-def1}), we obtain that
\[
G_t=n_t(t)E^2(t)+ n(t) E_t^2(t)\in L_\infty(0,T; \coH_{-2}).
\]
Using equation (\ref{Z_nd2}) from (\ref{est-M}) we have that
$ \|E_t^2(t)\|_{-2}\le C_M$ and thus
\[
 \|G(t)\| \le  C_M  \|n(t)\|
~~\mbox{and}~~
 \|G_t(t)\|_{-2}\le  C_M( \|n_t(t)\|+  \|n(t)\|_{2}).
\]
Therefore we can apply Lemma~\ref{le:schr-lin} with $a(t)=n^1(t)$
and $G(t)=n(t) E^2(t)$.
One can see that the corresponding function $W(t)$
admits the estimate
\[
a_0  \|E(t)\|_2^2 -C_M \|E(t)\|^2- \frac14 \|n(t)\|^2\le
W(t)\le C_M( \|E(t)\|_2^2+ \|n(t)\|^2).
\]
Thus from (\ref{en-Elin2}) we have that
\begin{multline*}
{}\quad
 a_0\|E(t)\|_2^2 \le  \frac14 \|n(t)\|^2+  C_M\left[ \|E(t)\|^2+ \|E(0)\|_2^2+ \|n(0)\|^2
\right] \\
+C_M \int_0^t\left[  \|n_t(\t)\|+ \|n(\t)\|_{2} +  \|E(\t)\|_2 \right]
 \|E(\t)\|_2  d\t.
\quad{}
\end{multline*}
By (\ref{en-E-lin1}) we have
\[
\|E(t)\|^2\le \|E(0)\|^2+C_M  \int_0^t \|n(\t)\|_{2}
 \|E(\t)\|  d\t.
\]
Therefore using (\ref{n-df}) and Gronwall's type argument
for the function
\[
\Phi(t)= \hf \cE_0(n_t(t),n(t))+a_0\|E(t)\|_2^2
\]
we obtain (\ref{lip}) and also uniqueness of weak solutions.
\par
We also note that the  continuity properties
in (\ref{Z-def2-0}) of weak solutions follow from
Lemmas \ref{le:n-lin} and \ref{le:schr-lin} applied to
equations (\ref{Z_nd1}) and (\ref{Z_nd2}) with "frozen"
nonlinearities.

\begin{remark}\label{re:weak-cont}
{\rm
We   note that
 the semiflow $S_t$ is also weakly continuous on $\cH$, i.e.,
 $S_{t}U_k\to S_tU$ weakly in $\cH$
for each $t\ge0$ when  $U_k\to U$ weakly in $\cH$ as $k\to\infty$.
Indeed, by (\ref{exist_9d+1})
the function $U_k(t)=S_tU_k$ satisfies the relation
\begin{equation}\label{ac-3}
\max_{[0,T]}\| U_k(t)\|_{\cH}\le C_T,\quad k=1,2\ldots
\end{equation}
Thus there exists a subsequence $\{k_j\}$ such that
$U_{k_j}(t)\to V(t)$ $*\!\mbox{-weakly~in}$
$L_\infty(0,T;\cH)$ as $j\to\infty$,
where $V(t)=(n_t;n;E)$ is a weak solution with initial data
$U$. By the uniqueness
$V(t)$ does not depend on the subsequence $\{k_j\}$ and, moreover,
 $V(t)=U(t)\equiv  S_tU$. Thus
$U_k(t)\to U(t)$  $*\!$-weakly in $L_\infty(0,T;\cH)$
as $k\to\infty$
and  by Aubin's embedding  theorem
(see \cite[Corollary 4]{sim})
$U_k(t)\to U(t)$ strongly in
$C(0,T;\cH_{-\si}\times\cH_{1-\si}\times\coH_{2-\si} )$
for every $\si>0$ when $k\to\infty$.
This and
(\ref{ac-3})  implies the conclusion.
}
\end{remark}

\subsection{Energy balance relation}
Now we prove (\ref{w-en2}).
\par
Since $n(t)$ is weak (variational) solution to
the linear problem (\ref{n-lin}) with
\[
F(t)\equiv -A (|E(t)|^2)+f\in C(0,T; H),
\]
using the multiplier
$A^{-1}n_t$ it is easy
to find that
\begin{align}\label{w-en2a}
V_0(n_t(t),n(t))
 +\alpha \int_0^t\|A^{-1/2}n_t\|^2 d\tau
 =
V_0(n_1,n_0)
-\int_0^t (n_t, |E|^2) d\tau.
\end{align}
We also note that $E(t)$ solves (\ref{schr-lin1}) with $a(t)=n(t)$
and $G(t)=g$. One can see that these $a$ and $G$ satisfy the hypotheses
of Lemma~\ref{le:schr-lin}.
Therefore (\ref{en-Elin2})  implies that
\begin{multline}\label{w-en2-5}
V_1(n(t),E(t))+2\gamma \int_0^t V_1(n,E) d\tau
 \\ =V_1(n_0,E_0)
+\int_0^t \left(  (n_t, |E|^2)+ 2 \gamma \Re(g,E)\right)d\tau.
\end{multline}
Taking the sum of (\ref{w-en2a})  and (\ref{w-en2-5}), we get
(\ref{w-en2}).

\subsection{Semi-strong solutions}
To establish the third statement of Theorem~\ref{th:wp}
we apply the second part of Lemma~\ref{le:schr-lin} with
 $a(t)=n(t)$ and $G(t)=g$.
Under the condition $E_0\in\coH_4$ this lemma imply (\ref{Z-def2-0sm2}).
\par
To obtain   (\ref{w-en})
for  semi-strong  solutions
we note that
 Lemma~\ref{le:n-lin} yields
 that
$(n_t; n)$
satisfies the energy  relation (\ref{lin-en-n0})
 with $F(t)=\Delta |E(t)|^2 +f$.
Therefore
\begin{equation}\label{Zen-n}
\cE_f(n_t(t), n(t))+\alpha\int_0^t\|n_t(\tau)\|^2d\tau=
\cE_f(n_1, n_0)
+\int_0^t (\Delta|E(\tau)|^2,n_t(\tau))d\tau,
\end{equation}
where $\cE_f(n_t,n)$ is given by (\ref{Zen-n1}).
Relation (\ref{Zen-E0lin}) implies
\begin{equation}\label{Zen-E}
\|E_t(t)\|^2+2\gamma \int_0^t\|E_t(\tau)\|^2d\tau
=
\|E_1\|^2
+2\Im \int_0^t (n_t(\tau) E(\tau),E_t(\tau))d\tau,
\end{equation}
where $E_1\in\cH$ is given by (\ref{Zen-E1}).
Since
\[
(\Delta|E|^2,n_t)=2(n_t, |\g E|^2)+2\Re (n_tE,\Delta E),
\]
relation (\ref{w-en}) follows from (\ref{Zen-n}) and (\ref{Zen-E}).
\par
To continue with properties of semi-strong solutions
we  note  that directly from (\ref{Z_nd2}) we have that
\begin{equation}\label{et-de}
\|iE_t(t)-(A+h^2A^2)E(t)\|\le
 \|g\|+ C\left(1+\|n(t)\|_2\right) \|E(t)\|.
\end{equation}
Hence
one can see from (\ref{w-en}) and (\ref{exist_9d+1})
via Gronwall's type argument that
\begin{equation}
\label{est-h*}
\|n_t(t)\|^2+\|n(t)\|_2^2+ \|E_t(t)\|^2+\|E(t)\|_4^2\le C_{T,R}
(1+\|E_1\|^2)
 \end{equation}
for  $t\in [0,T]$  and for $Y_0=(n_1;n_0;E_0)\in \cH_*$ such that
$\|Y_0\|_\cH\le R$.
\par
Now to prove the Lipschitz property in (\ref{lip-h*})
we use the fact that  the difference
(\ref{dif-def}) of two solutions satisfies (\ref{Z_nd1-df})
and (\ref{Z_nd2-df}).
Applying Lemma~\ref{le:schr-lin} to equation (\ref{Z_nd2-df})
from (\ref{Zen-E0lin}) we obtain
\begin{equation*}
\d\|E_t\|^2\le C\left(
\|n^1_t\| \|E\|_2 +\|n_t\| \|E^2\|_2 +\|n \|_2 \|E_t^2\|\right)  \|E_t\|.
 \end{equation*}
By (\ref{est-h*}) applied to the both solutions $(n^i;E^i)$
 this yields
\begin{multline*}
\|E^1_t(t)-E^2_t(t)\|^2\le \|E^1_t(0)-E^2_t(0)\|^2
+\int_0^t
 \|E^1_t(\t)-E^2_t(\t)\|^2 d\t
\\ + C\int_0^t
\left( \|E^1(\t)-E^2(\t)\|_2^2+
\|n^1_t(\t)-n^2_t(\t)\|^2 +\|n^1(\t)-n^2(\t) \|^2_2\|\right)d\t,
\end{multline*}
where the constant $C$ depends on $R$ and $T$.
Therefore using (\ref{lip}) we obtain the Lipschitz property in (\ref{lip-h*}).

\section{Global attractor}
\label{sec:glob-at}
The main goal in this section is to prove
Theorem \ref{t:attr} which states the existence of finite dimensional
global attractor in the case when the dissipation parameters
$\alpha$ and $\ga$ are positive.
\par
According to the  general theory of dissipative systems
(see, e.g., \cite{BV92,Chu99,Lad91,Temam}) to prove the existence
of a compact global attractor we need to establish dissipativity
and asymptotic compactness of the corresponding dynamical system.
\par
We do this according to the following plan.
We first establish several dissipativity properties
of the systems generated by (\ref{Z_nd1})--(\ref{Z_nd3}).
Then in Section~\ref{Sec3.2} using the  splitting method
we prove that the system possesses a compact attracting set
which is bounded in a partially smoother space $\cH_*$. This implies
the existence of a global attractor $\frA$.
In Section~\ref{Sec3.3} we consider the restriction of
the system $(\cH,S_t)$ on the space $\cH_*$
and prove that $(\cH_*,S_t)$   admits a stabilizability
estimate (see Proposition~\ref{pr:q-stab}). As in \cite{ChuLas} and \cite[Chapter 7,8]{cl-book}
this allows us  to prove the existence of finite dimensional attractor
$\frA_*$ for $(\cH_*,S_t)$. Then we show that $\frA=\frA_*$.
By Proposition~\ref{t:attr*}
concerning the attractor $\frA_*$
 this implies the statements 1-3 of
Theorem~\ref{t:attr}. In Sections \ref{Sec3.4} and \ref{Sec3.5}
using some ideas developed in \cite{GT87} and
\cite{ChuLas}
we establish the smoothness properties of the
attractor.

\subsection{Dissipativity}\label{Sec3.1}
We first note that
by Theorem~\ref{th:wp} we have the following dissipativity property
in the variable $E$:
\begin{equation}\label{a2}
\limsup_{t\to\infty}\left[\sup\left\{\|E(t)\|\; :\; (n_1;n_0;E)\in B\right\}
\right]\le \gamma^{-1}\|g\|
\end{equation}
for any bounded set $B$ in $\cH$. We use it to obtain dissipativity
in $\cH$ in two steps.
We first  prove  the following assertion on
 dissipativity in
$H_{-1}\times H_1\times \coH_2$.
\begin{lemma}\label{l:dis2}
Let $\a,\gamma>0$.
Then there exists  a
constant $R_0>0$  such that
\begin{equation}\label{a3}
\limsup_{t\to\infty}\left[\sup\left\{\|n_t(t)\|^2_{-1}+\|n(t)\|_1^2
+\| E(t)\|_2^2\; :\; (n_1;n_0;E)\in B\right\}\right]\le R_0
\end{equation}
for any bounded set $B$ in $\cH$.
\end{lemma}
\begin{proof}
To prove (\ref{a3}) we consider the following Lyapunov type function
\begin{equation*}
\cW(n_t,n,E)=V(n_t,n,E)+\eps\left[ (n,A^{-1}n_t)+
\frac{\alpha}2 \|A^{-1/2}n\|^2
\right],
\end{equation*}
where  $V(n_t,n,E)= V_0(n_t,n)+ V_1(n,E)$ with
 $V_0(n_t,n)$
 defined  by (\ref{Zen-n1-2})  and $V_1(n,E)$ by (\ref{r-en-2}).
The parameter $\eps>0$ will be chosen later.
\par
Let $B$ be a bounded set in $\cH$. By (\ref{a2}) there exists
 $t_B\ge 0$  such that
\begin{equation}\label{a3-B}
\|E(t)\| \le \rho\equiv 1+\gamma^{-1}\|g\|~~~\mbox{for all}~~
t\ge t_B.
\end{equation}
Then it follows from (\ref{v-v+}) with $K=\rho$ that
\begin{equation}\label{a6-1}
\cW(n_t,n,E)\ge \left(a_0-\frac{\eps}{2\alpha}\right)
\|A^{-1/2}n_t\|^2+
a_0 V_+(n_t,n,E) - a_1,~~~t\ge t_B,
\end{equation}
for some  $a_i>0$,
where $ V_+(n_t,n,E)$ is given by (\ref{exist_5b+}).
We also have that
\begin{eqnarray*}
\lefteqn{
 \frac{d}{dt}\left[ (n,A^{-1}n_t)+
\frac{\alpha}2 \|A^{-1/2}n\|^2\right]
}\nonumber  \\ & &
\qquad = \|A^{-1/2}n_t\|^2-\|n\|^2- h^2\|n\|^2_1-(n, |E|^2)+(f,A^{-1}n)
\end{eqnarray*}
for almost all $t\ge 0$.
Therefore,
from (\ref{w-en2})  we have that
\begin{align*}
 \frac{d}{dt}\cW(n_t,n,E) &+ \eps\cW(n_t,n,E)
\nonumber
\\
  = &
-(2\gamma-\eps)(\|E\|_1^2+h^2\|E\|_2^2)
-2\gamma(n, |E|^2)-2(\gamma-\eps)\Re(g,E) \nonumber
  \\
  & - (\alpha-\fracd{3\eps}{2})\|A^{-1/2}n_t\|^2
  -\fracd{\eps}{2} (\|n\|^2+h^2\|n\|_1^2) \nonumber
\\  &
+\eps^2\left[ (n,A^{-1}n_t)+
\frac{\alpha}2 \|A^{-1/2}n\|^2
\right].
\end{align*}
By (\ref{a3-B}) from (\ref{exist_8a})  we obtain that
\[
|(n, |E|^2)| \le  \eta (\|n\|^2+ \| E\|_2^2)+  C_{\eta}(\rho), ~~~
t\ge t_B.
\]
Therefore, taking into account the relation
\[
\left(n,A^{-1}n_t\right)+ \frac{\alpha}2\| A^{-1/2} n\|^2 \le
 \frac{\alpha}{\lambda_1}\|n\|^2
+\fracd{1}{2\alpha}\|n_t\|_{-1}^2
\]
and an obvious estimate for the linear term $(g,E)$,
 we get that
\[
 \frac{d}{dt}\cW(n_t,n,E)+\eps \cW(n_t,n,E)
 \le C(g,f,\rho)~~~\mbox{for $t\ge t_B$,}
\]
where $\eps>0$ is small enough.
 Therefore for $\cW(t)\equiv\cW(n_t(t),n(t),E(t))$
we obtain the estimate
\[
\cW(t)\le \cW(t_B)\exp\left\{-\eps(t-t_B)\right\}
+C_\eps(g,f,\rho)
~~~\mbox{for all $t\ge t_B$.}
\]
Consequently relation (\ref{a3}) follows from (\ref{a6-1}).
\end{proof}
Now we can establish dissipativity in $\cH$.
\begin{lemma}\label{l:dis2-h}
Let $\a,\gamma>0$.
Then there exists  a
constant $\tilde{R}>0$  such that
\begin{equation}\label{a3-h}
\limsup_{t\to\infty}\left[\sup\left\{\|n_t(t)\|^2+\|n(t)\|_2^2
+\| E(t)\|_2^2\; :\; (n_1;n_0;E)\in B\right\}\right]\le \tilde{R}
\end{equation}
for any bounded set $B$ in $\cH$, i.e., the system $(\cH,S_t)$ is dissipative.
\end{lemma}
\begin{proof}
To prove (\ref{a3-h})
 we use the same type calculations as in the previous lemma and
consider the following Lyapunov type function
\[
\cW_*(n_t,n)=\cE_0(n_t,n)+\eps\left[ (n,n_t)+
\frac{\alpha}2 \|n\|^2
\right],
\]
where $\cE_0(n_t,n)$
is given by (\ref{lin-en-n1}).
The parameter $\eps>0$ will be chosen later.
\par
By Lemma~\ref{l:dis2} for every
bounded set  $B$ in $\cH$ there exists  $t_B\ge 0$  such that
\begin{equation*}
\|n_t(t)\|_{-1}^2+\|n(t)\|_1^2
+\| E(t)\|_2^2\le 1+R_0
~~~\mbox{for all}~~
t\ge t_B.
\end{equation*}
Therefore
applying  Lemma~\ref{le:n-lin} to equation (\ref{n-lin})
with $F(t)=\Delta|E(t)|^2 +f$
we get from (\ref{lin-en-n0}) that
\[
\d \cE_0(n_t,n)\le - \frac{\a}2 \|n_t\|^2 +C_{R_0} ~~~\mbox{for}~~
t\ge t_B.
\]
In a similar way
one can also see that
\[
\d \left[ (n,n_t)+
\frac{\alpha}2 \|n\|^2
\right]\le \|n_t\|^2-\hf \left[ \|n\|_1^2 +h^2\|n\|^2_2\right] +C_{R_0},
~~~
t\ge t_B.
\]
Therefore choosing $\eps$ small enough we obtain
 that
\[
 \frac{d}{dt}\cW_*(n_t,n)+\eps \cW_*(n_t,n)
 \le C_{R_0}~~~\mbox{for $t\ge t_B$.}
\]
This and also Lemma~\ref{l:dis2}  imply (\ref{a3-h}).
\end{proof}
Now we  prove
 dissipativity of semi-strong solutions in their phase  space
 $\cH_*=H\times H_2\times \coH_4$. We need this to
prove the existence of an attractor in $\cH_*$.
\begin{lemma}\label{l:dis3}
Let $\a,\gamma>0$.
Then there exists  a
constant $R_*>0$  such that
\begin{equation}\label{a3-h*}
\limsup_{t\to\infty}\left[\sup\left\{\|n_t(t)\|^2+\|n(t)\|_2^2
+\| E(t)\|_4^2\; :\; (n_1;n_0;E)\in B\right\}\right]\le R_*
\end{equation}
for any bounded set $B$ in $\cH_*$. Thus
the dynamical system
 $(\cH_*,S_t)$ generated by problem   (\ref{Z_nd1}) and (\ref{Z_nd2})
is dissipative (in the topology of the space $\cH_*$).
\end{lemma}
\begin{proof}
On the trajectories $(n_t;n;E)$ we consider the following  function
\begin{equation}\label{b1-a}
W(t)\equiv W(n_t;n;E_t)=\cE_f(n_t,n)+ \|E_t\|^2
+\eps\left[ (n,n_t)+
\frac{\alpha}2 \|n\|^2\right],
\end{equation}
where $\cE_f(n_t,n)$ is given by (\ref{Zen-n1}).
The parameter $0<\eps\le\alpha/2$ will be chosen later.
It is clear that
\begin{equation}\label{b2}
c_0 W_0(t)-c_1\|f\|^2
\le W(t)\le c_2 W_0(t)+c_3\|f\|^2
\end{equation}
with some positive constants $c_i$,
where
\[
W_0(t)\equiv \cE_0(n_t,n)+ \|E_t\|^2 =
\frac12\left[ \|n_t\|^2+\| n\|_1^2+ \|n\|_2^2\right] +\| E_t\|^2.
\]
Since
\[
\d\left[ (n,n_t)+
\frac{\alpha}2 \|n\|^2\right]
= \|n_t\|^2-\| n\|_1^2-h^2\|n\|^2_2+(n, \Delta |E|^2+f),
\]
from (\ref{w-en}) we obtain that
\begin{equation}\label{b3}
\d W(t) = -(\alpha-\eps)\|n_t\|^2-\eps(\| n\|_1^2+h^2\|n\|^2_2)-2\gamma \| E_t\|^2+
Q_\eps(n_t;n;E)
\end{equation}
for almost all $t\ge 0$,
where
\[
Q_\eps(n_t;n;E)=2(n_t,|\g E|^2)+2\Re(n_tE, iE_t +\Delta E )+
\eps (n, \Delta |E|^2+f).
\]
By Lemma~\ref{l:dis2-h} there exists
 $t_{B}\ge0$  such that
\begin{equation}\label{b2-diss-0}
\|n_t(t)\|^2+\|n(t)\|_2^2+\|E(t)\|_2^2\le 1+ \tilde{R}
\end{equation}
for $t\ge t_{B}$.
Therefore after simple calculations it is easy to see that
\[
|Q_\eps(n_t;n;E)|\le C(\tilde{R})(1+\|E_t\|)\le \ga
 \|E_t\|^2 + C(\tilde{R}),
\quad t\ge t_{B}.
\]
Consequently from (\ref{b2}) and (\ref{b3})
we have that
\[
\d W(t)\le -\om W(t)+C,\quad t\ge t_{B},
\]
for some $\om>0$, which yields

\begin{equation}\label{b2-diss}
W_0(t)\le  c_1 W_0(t_B) e^{-\om (t-t_B)} +c_2, \quad t\ge t_{B},
\end{equation}
with some positive constants $c_i$.
It follows from (\ref{et-de}) and (\ref{b2-diss-0}) that
\[
\| E\|_4\le h^{-2}\|E_t\|+ C_1(\tilde{R})~~~\mbox{
for $t\ge t_{B}$.}
\]
Therefore (\ref{b2-diss}) implies (\ref{a3-h*}) and hence $(\cH_*,S_t)$
is dissipative.
\end{proof}

\subsection{Asymptotic compactness  in $\cH$}
\label{Sec3.2}
We recall the following definition (see, e.g., \cite{Lad91,Temam}
or \cite{Chu99}).
\begin{definition}\label{de:as-comp}
{\rm
A dynamical system $(X,S_t)$  on a complete metric
space $X$ is said to be {\em asymptotically compact} if
 for any bounded set $B$ from $X$
there exists a compact set $K$ in  $X$ such that
$\sup
\left\{{\rm dist}_X (S_tx, K)\, :\,
 x\in B  \right\}\to 0$
as $t\to\infty$.
}
\end{definition}
The following assertion
shows that  $(\cH,S_t)$ is  asymptotically compact
in  $\cH$.

\begin{lemma}\label{le:ac-H}
Let $0<\si<1/2$ and  $\cH_\si= H_\si\times  H_{2+\si}\times \coH_4$.
There exist a ball
$\sB_\si(R)=\left\{ \|U\|_{\cH_\si}\le R  \right\}$ in $\cH_\si$
and a number $\delta>0$ such that
\begin{equation}\label{ac-H-si}
\sup\left\{{\rm dist}_\cH (S_tU,\sB_\si(R))
\, :\; U\in B\right\} \le C_B e^{-\delta t},~~t>0,~~
\end{equation}
for any bounded set $B\subset\cH$.
This means that $(\cH,S_t)$ is asymptotically compact system
because $\cH_\si$ is compactly embedded in $\cH$.
\end{lemma}
\begin{proof}
By Lemma \ref{l:dis2-h} we can assume that
$S_tU=(n_t(t);n(t);E(t))$ possesses the property
(\ref{b2-diss-0}) for {\em all} $t>0$ and $U\in B$.
\par
We use the following version of the splitting method.
We first split the $E$-component as
$E(t)=E^s(t)+E^c(t)$, where   $E^s(t)$ solves the problem
\begin{equation*}
    i E^s_t-(A + h^2 A^2) E^s- n E^s+i\gamma E^s =0,~~~  E^s|_{t=0}=E_0,
\end{equation*}
and $E^c$ is solution to
\begin{equation}
\label{Z_nd2-comp}
    i E^c_t-(A + h^2 A^2) E^c- n E^c+i\gamma E^c =g,~~~  E^c|_{t=0}=0.
\end{equation}
It follows from Lemma~\ref{le:schr-lin} that
\begin{equation}
\label{E-st}
   \|E^s(t)\|^2\le e^{-2\ga t}\|E_0\|^2, ~~~ t\ge 0.
\end{equation}
and also
\begin{equation}\label{E-st2}
\d W^s(t)+2\gamma  W^s(t)
\le
 |(n_t(t), |E^s(t)|^2)|,
\end{equation}
where
$W^s(t)= \|E^s(t)\|_1^2+h^2\|E^s(t)\|_2^2 + (n(t),|E^s(t)|^2)$.
Similar to (\ref{exist_8a}) using (\ref{b2-diss-0}) we have that
\[
 |(n_t, |E^s|^2)|\le \ga h^2\|E^s\|^2_2+
 C_{\tilde{R}} \|E^s\|^2
~~~
\mbox{and}
~~~
 |(n, |E^s|^2)|\le  C_{\tilde{R}} \|E^s\|^2.
\]
Therefore (\ref{E-st}) and (\ref{E-st2})  imply that
 \begin{equation}
\label{E-st-main}
   \|E^s(t)\|_2^2\le  C_{\tilde{R}}  e^{-\ga t}
~~~\mbox{for all}~~ t\ge 0.
\end{equation}
Now we consider (\ref{Z_nd2-comp}). It follows
from Lemma~\ref{le:schr-lin} and from (\ref{b2-diss-0})
that
 \begin{equation}
\label{E-com}
\d\|E_t^c(t)\|^2 +2\gamma  \|E^c_t(t)\|^2
\le
2  |(n_t(t) E^c(t), E^c_t(t))|\le \gamma  \|E_t^c(t)\|^2 +
C_{\tilde{R}}  \|E^c(t)\|_2^2.
\end{equation}
It follows from (\ref{b2-diss-0})  and (\ref{E-st-main})
that $ \|E^c(t)\|_2^2= \|E(t)-E^s(t)\|_2^2\le C_{\tilde{R}}$
for all $t\ge 0$. Thus we have from (\ref{E-com}) that
$\|E^c_t(t)\|^2\le  C_{\tilde{R}}$ for all $t\ge 0$.
Since
\[
 \|i E^c_t-(A + h^2 A^2) E^c\|\le \|n E^c-i\gamma E^c+g\|\le  C_{\tilde{R}}.
\]
we conclude that
\begin{equation}\label{E-com-main}
\|E^c(t)\|_4^2\le  C_{\tilde{R}}~~~\mbox{for all}~~t\ge 0.
\end{equation}
\par
Now we switch on the  $n$-component.
Let $Y(t)=(n_t(t); n(t))$. Using the constant variation formula we have
that
\begin{equation}\label{n-split}
Y(t)=  I^t +K^t\equiv  I^t +\int_0^t T_{t-\t}
\big(\Delta|E^c(\t)|^2;0\big)d\t +  (0; \cA_h^{-1}f), ~~~ t\ge s,
\end{equation}
where   $\cA_h=A+h^2A^2$, $T_t$ is $C_0$-semigroup generated by
(\ref{n-lin}) with $F\equiv 0$, and
\begin{equation*}
 I^t= T_{t}\big[ Y(0)- (0; \cA_h^{-1}f) \big]+\!\int_0^t \! T_{t-\t}
\big(\Delta\big[|E^s(\t)|^2+2\Re E^s(\t)E^c(\t)\big];0\big)d\t.
\end{equation*}
Using (\ref{T-exp-st}) with $\si=0$
and also (\ref{b2-diss-0}), (\ref{E-st-main}) and (\ref{E-com-main})
we have that
\begin{align}\label{I-stab}
\| I^t\|_{H\times H_2}& \le  C_{\tilde{R}}   e^{-\om t} + C_{\tilde{R}}  \!
\int_0^t \!e^{-\om(t-\t)}
\left[\|E^s(\t)\|_2^2+ \|E^s(\t)\|_2\|E^c(\t)\|_2\right]d\t
\nonumber
  \\ & \le C_{\tilde{R}}
e^{-\delta t} ,~~~ t>0, ~~\mbox{for some}~~\delta>0.
\end{align}
In a similar way, for $\si<1/2$ we have
$\Delta|E^c(\t)|^2\in \sD(A^\si)$ and thus
by \eqref{E-com-main} we have that
  \begin{align}\label{K-comp}
\| K^t\|_{H_\si\times H_{2+\si}}& \le M \int_0^t e^{-\om(t-\t)}
\|E^c(\t)\|^2_{4} d\t+ C\|f\|_{-2+\si} \le C_{\tilde{R}}
\end{align}
for all $t\ge  0$.
Let $B_\si(\varrho) =\{ w\in H_\si\times H_{2+\si}:
\|w\|_{H_\si\times H_{2+\si}}\le \varrho\}$.
Then
it follows from (\ref{n-split})--(\ref{K-comp})
that
there exists $\varrho>0$ large enough such that
\begin{equation}\label{ac-H-1}
{\rm dist}_{H\times H_2}\big(Y(t), B_\si(\varrho)\big)\le
 C_{\tilde{R}}
 e^{-\eta t},~~~ t\ge 0,
\end{equation}
for some $\eta>0$.
In a similar way from (\ref{E-st-main}) and (\ref{E-com-main})
we have that there exists a ball
 $D(\varrho_*) =\{ u \in \coH_4:
\|u\|_{4}\le \varrho_*\}$
 such that
\begin{equation}\label{ac-H-2}
{\rm dist}_{\coH_2}\big(E(t), D(\varrho_*)\big) C_{\tilde{R}}
 e^{-\ga t/2},~~~ t\ge 0.
\end{equation}
Thus  (\ref{ac-H-si}) follows from (\ref{ac-H-1}) and
(\ref{ac-H-2}).
The proof of Lemma~\ref{le:ac-H} is complete.
\end{proof}
\par
The existence of the compact global attractor $\frA$
for $(\cH,S_t)$ now follows
by the standard results (see, e.g., \cite{Temam} or \cite{Chu99})
from Lemmas~\ref{l:dis2-h} and~\ref{le:ac-H}.
Moreover by (\ref{ac-H-si}) $\frA$ is a bounded set in the space
$\cH_\si= H_\si\times  H_{2+\si}\times \coH_4$
for every $0<\si<1/2$, in particular $\frA$    is bounded
in $\cH_*$.
\smallskip\par
To prove other statements in Theorem~\ref{t:attr} we first
consider long-time dynamics of semi-strong solutions.

\subsection{Semi-strong attractor}\label{Sec3.3}
In this section we prove in Proposition~\ref{t:attr*}
the existence of a global attractor $\frA_*$
of the evolution semigroup   $S_t$
in the space $\cH_*= H\times  H_{2}\times \coH_4$.
Then we show that $\frA_*$ coincides with the attractor $\frA$.
This allows us to establish the statements 1-3 in Theorem~\ref{t:attr}.
Proposition~\ref{t:attr*} provides also some steps in
the proof of smoothness properties of the attractor $\frA$.

\begin{proposition}\label{t:attr*}
 Assume that $\alpha,\ga >0$ and $(f;g) \in H\times \coH$.
Then
the dynamical
system $(\cH_*,S_t)$ generated by   (\ref{Z_nd1})--(\ref{Z_nd3})
in the space $\cH_*$
possesses a compact global attractor $\frA_*$.
Moreover:
\begin{enumerate}
    \item[{\bf (1)}] This attractor $\frA_*$ has finite fractal dimension.
  \item[{\bf (2)}]  $\frA_*$ is a bounded set in
$H_2\times H_4\times\coH_4$ and for any trajectory
\[
U(t)=(n_t(t);n(t);E(t)),~~~ -\infty<t<+\infty,
\]
from the attractor we have that
\[
(n_{tt};n_t;n)\in C(\R; H\times H_2\times H_4),~~~
(E_t;E)\in C^1(\R; \coH\times\coH_4)
\]
 and there exists $R>0$ (independent of $U$) such that
\begin{equation}\label{U-atr-bnd*}
\sup_{t\in\R}\left\{\|n_{tt}(t)\|^2+\|n_{t}(t)\|^2_2+
\|n(t)\|^2_4+\|E_{tt}(t)\|^2 +\|E_{t}(t)\|^2_4  \right\}
\le R.
\end{equation}
\end{enumerate}
\end{proposition}

To prove this proposition we use recently
developed approach based on stabilizability estimates
(see, e.g., \cite{ChuLas} and \cite[Chap.7,8]{cl-book}).

\begin{proposition}[{\bf Stabilizatility estimate}]\label{pr:q-stab}
Let $(n^i(t);E^i(t))$, $i=1,2$, be two semi-strong solutions
with different initial data $U^i=(n_1;n_0;E_0)$
such that
\begin{equation*}
\|n^i_t(t)\|^2+\|n^i(t)\|_2^2+\|E^i_t(t)\|^2+\|E^i(t)\|_4^2\le R,~~~ t\ge 0,
\end{equation*}
for some $R>0$.
 Then the    difference
\begin{equation*}
U(t)\equiv (n_t(t);n(t);E(t))=((n^1_t(t)-n^2_t(t);n^1(t)-n^2(t);E^1(t)-E^2(t))
\end{equation*}
satisfies the relation
\begin{equation}\label{stab-est}
\|U(t)\|_{\cH_*}^2\le C_Re^{-\varkappa t}\|U(0)\|_{\cH_*}^2
+C_R\left[ \|E(t)\|^2+\int_0^t e^{-\varkappa (t-\t)}\|E(\t)\|^2d\t\right],
\end{equation}
where $C_R$ and  $\varkappa$ are positive constants.
\end{proposition}

\begin{proof}
Since $n(t)$ solves (\ref{Z_nd1-df}),  using the energy relation in
Lemma~\ref{le:n-lin} we have  that
there exist $\delta,\eta>0$ such that
 the function
\[
W(t)=\frac12 \left[\|n_t\|^2 +\|n\|_1^2+h^2\|n\|_2^2\right]
+\delta \left[ (n,n_t) +\frac{\a}2\|n\|^2\right]
\]
satisfies relations
$
a_0\left[\|n_t\|^2 +\|n\|_2^2\right]\le W(t)
\le a_1\left[\|n_t\|^2 +\|n\|_2^2\right]
$ and
\[
\d W(t)+\eta W(t)\le -a_2  \left[\|n_t\|^2 +\|n\|_2^2\right]
+\eps \|E\|^2_4 +C_{R,\eps}\|E\|^2
\]
for any $\eps>0$, where $a_i>0$ are constants.
Since $E$ solves (\ref{Z_nd2-df}), we have from (\ref{Zen-E0lin}) that
\begin{align*}
\d\|E_t(t)\|^2 +\ga   \|E_t(t)\|^2 &\le
C\| n^1_t E+n_t E^2+ n E_t^2\|^2 \\
&  \le  \eps \|E\|^2_4 +C_{R,\eps}\|E\|^2 +C_R\left[
\| n_t\|^2+ \|n\|_2^2\right].
\end{align*}
Let $W_\mu(t)=W(t)+\mu \|E_t(t)\|^2$.
Choosing $\mu>0$ small enough
we obtain that
\begin{equation}\label{W-n-est-2}
\d W_\mu(t)+ c_0 W_\mu (t)\le
\eps \|E(t)\|^2_4 +C_{R,\eps}\|E(t)\|^2,
\end{equation}
where $c_0=\min\{ \eta,\ga\}$.
One can see from (\ref{Z_nd2-df}) that
\[
\|iE_t-(A+h^2A^2)E\|=\|i\gamma E-n^1 E -n E^2\|\le C_R(\|E\|+\|n\|).
\]
Thus
\[
\|E\|_4\le h^{-2}\|(A+h^2A^2)E\|\le h^{-2} \|E_t\|+  C_R(\|E\|+\|n\|).
\]
Therefore the estimate in (\ref{stab-est}) follows from
(\ref{W-n-est-2}).
\end{proof}
Now we can apply Ceron-Copes type criteria (see \cite[Corollary 2.7]{ChuLas})
and also the dissipativity stated in Lemma~\ref{l:dis3} to guarantee
the existence of a  global compact attractor $\frA_*$ in $\cH_*$.
\par
To prove  finiteness of fractal dimension of $\frA_*$ we use
the method based on the idea of short trajectories
due to J.M\'{a}lek and J.Ne\v{c}as
(see \cite{malek-ne} and also \cite{malek} and \cite{ChuLas_JDDE_2004})
 and stabilizability estimates.
\par
 For some $T\ge 1$ which we specify later
 we consider  the space
\[
W_T=\left\{E\in  L_2(0,T;  \coH_4)\, :\; E_t\in
 L_2(0,T;  \coH)   \right\}
\]
with the norm
\[
|u|_{W_T}^2=\int_0^T\left[ \|E_t(t)\|^2+ \|E(t)\|_{4}^2\right] dt.
\]
Let $\cW_T=\cH_*\times W_T$ with the norm
$|(Y; G)|^2_{\cW_T} =\|Y\|^2_{\cH_*}+|G|^2_{W_T}$.
We denote by $\frA_T$ the set in $\cW_T$ of the form
\[
\frA_T=\{ W=(Y; E(\cdot)) : Y=(n_1;n_0;E_0)\in \frA_* \},
\]
where
$(n;E)$ is a solutions
 on the interval $[0,T]$
with initial data  $(n_1;n_0;E_0)$ from the attractor  $\frA_*$.
It is clear that
${\frA}_T$ is a closed bounded set in $\cW_T$.
On $\frA_T$ we define the shift operator $V$
by the formula
\[
V\, :\; {\frA}_T\mapsto {\frA}_T, ~~~ V(Y,E)=
(S_TY; E(T+t),~  t\in [0,T]).
\]
It is  clear that ${\frA}_T$ is strictly invariant \wrt $V$, i.e.
$V{\frA}_T={\frA}_T$.
\par
In further calculations we use the same
idea as in \cite{ChuLas_JDDE_2004},
see also \cite{ChuLas,cl-book}.
\par
It follows   from the Lipschitz estimate in (\ref{lip-h*}) that
\begin{equation*}
|V U_1-VU_2|_{\cW_T}\le C_T | U_1- U_2|_{\cW_T},~~~
U_1,U_2\in\frA_T.
\end{equation*}
From  Proposition~\ref{pr:q-stab}
we also have that
\begin{equation}\label{qst-w-est}
|V U_1-VU_2|^2_{\cW_T}\le q_{T} | U_1- U_2|^2_{\cW_T} +C_T
\left[ n_T^2(U_1-U_2) +n_T^2(VU_1-VU_2)\right]
\end{equation}
for every $ U_1,U_2\in {\frA}_T$, where
$q_T= C e^{-\varkappa T}$ and
 the seminorm $n_T(U)$ has the form
\[
n^2_T(U) \equiv \max_{t\in [0,T]} \| E(t)\|^2
~~\mbox{for}~~ U= (Y;E(\cdot))\in  \cW_T.
\]
One can see that this seminorm is compact on $W_T$.
Therefore  we can choose $T\ge 1$ such that $q_T<1$ in (\ref{qst-w-est})
 and apply Theorem 2.15\cite{ChuLas}
to conclude that   $\frA_T$   has a finite fractal dimension in $\cW_T$.
This implies the finiteness of the fractal dimension of
$\frA_*$ in $\cH_*$.
\par
Now we prove the smoothness result in (\ref{U-atr-bnd*})
using the same method as \cite{ChuLas} and \cite[Chapter 7,8]{cl-book}.
For this we apply stabilizability estimate  (\ref{stab-est}) to
the trajectory $U(t)$ and to its shift
$U_\delta =\{U(\delta+t) : t\in\R \}$ with starting time $s$
instead of $0$. Since
 $\|E(t+\delta)-E(t)\|\le C_R\delta^2$ for every $t\in \R$,
we have that
\[
\|U(t+\delta)-U(t)\|_{\cH_*}^2\le C_Re^{-\varkappa (t-s)}
\|U(s+\delta)- U(s)\|_{\cH_*}^2+ C_R\delta^2.
\]
for all $s\le t$.
Therefore  after the limit transition  $s\to -\infty$  we obtain
\[
\frac1{\delta^2}\left[\|n_t(t+\delta)- n_t(t)\|^2+\||n(t+\delta)- n(t)\|_2^2+
\||E(t+\delta)- E(t)\|_4^2\right]\le C_R
\]
for every $ t\in \R$. In the limit $\delta\to 0$ this gives
the conclusion in (\ref{U-atr-bnd*})
and completes the proof of Proposition~\ref{t:attr*}.
\smallskip\par \noindent
{\bf Proof of Theorem~\ref{t:attr}(1-3):}
It is clear that the attractor $\frA_*$ is included in
the attractor $\frA$ for the system $(\cH,S_t)$.
Since $\frA$ is an strictly invariant bounded set in $\cH_*$,
we also have that $\frA\subset \frA_*$ and thus
 $\frA= \frA_*$. Therefore the statements (1)-(3) in Theorem~\ref{t:attr}
follows from Proposition~\ref{t:attr*}.
\smallskip\par
Thus it remains to establish time and spatial smoothness
stated in Theorem~\ref{t:attr}(4,5).

\subsection{Smoothness in time variable}
\label{Sec3.4}
To prove Theorem~\ref{t:attr}(4) it is convenient to
apply the induction method in the form presented in the following
assertion.

\begin{proposition}\label{pr:t-smo}
Let  $U(t)=(n_t(t);n(t);E(t))$, $t\in\R$, be a trajectory
from the attractor $\frA$. Then
\begin{equation}\label{sm-1}
(n_{tt};n_t;n;E_{tt};E_t)\in C^{k-1}(\R; H\times H_2\times H_4\times \coH\times\coH_4)~~~\mbox{for $k=1,2,\ldots$}
\end{equation}
 and there exists a non-decreasing  sequence $\{R_k\}$ such that
\begin{multline}\label{sm-2}
\|n^{(k+1)}(t)\|^2+\|n^{(k)}(t)\|^2_2+
\|n^{(k-1)}(t)\|^2_4\\ +\|E^{(k+1)}(t)\|^2 +\|E^{(k)}(t)\|^2_4 +
\|E^{(k-1)}(t)\|^2_4
\le R_k
\end{multline}
for all $t\in\R$ and for every  $k=1,2,\ldots$,
where $n^{(m)}(t)$ and $E^{(m)}(t)$ denote time derivatives
of the order $m$.
\end{proposition}
\begin{proof}
We use the induction in $k$. By Proposition~\ref{t:attr*} the statement
of Proposition~\ref{pr:t-smo} is valid for $k=1$. Assume that
(\ref{sm-1}) and (\ref{sm-2}) hold for {\em all} $1\le k\le m-1$
with $m\ge 2$. As in \cite[Theorem 9.5.5]{cl-book} we
use the idea of higher order
stabilizability estimates inspired by the approach
presented in \cite{GT87}.

\par
Let $\widetilde{n}(t)=n^{(m-1)}(t)$ and $\widetilde{E}(t)=E^{(m-1)}(t)$.
By the induction hypothesis
 \[
(\wn_{t};\wn;\wE_t;\wE)\in C
(\R; H\times H_2\times  \coH\times\coH_4)
\]
and the following equations
\begin{align}
\label{w-nd1}
   & \wn_{tt}+ \left(A+ h^2 A^2\right)\wn+\alpha \wn_t  =F^{m-1}(E;t),
 \\
 &
\label{w-nd2}
    i \wE_t-(A + h^2 A^2) \wE -n(t)\wE +i\gamma \wE =G^{m-1}(n,E;t),
\end{align}
are satisfied,    where
\begin{equation}
\label{F-m}
F^{m-1}(E;t)=\Re \sum_{j=0}^{m-1} C^j_{m-1}\Delta\Big[ E^{(j)}(t) \overline{E^{(m-1-j)}}(t)\Big]
\end{equation}
(the bar over $E^{(m-1-j)}$ means the complex conjugations)  and
\begin{equation}
\label{G-m}
G^{m-1}(n,E;t)= \sum_{j=1}^{m-1} C^j_{m-1} n^{(j)}(t) E^{(m-1-j)}(t).
\end{equation}
Here $C^j_k$ denotes the binomial coefficients.
\begin{lemma}\label{le:FG}
Let
(\ref{sm-1}) and (\ref{sm-2}) be valid  for {\em all} $1\le k\le m-1$
with $m\ge 2$ for two trajectories
 $U_i(t)=(n_{it}(t);n_i(t);E_i(t))$, $t\in\R$, $i=1,2$,
 from the attractor.
Let $n_*(t)=n_1(t)-n_2(t)$ and  $E_*(t)=E_1(t)-E_2(t)$.
Then there exists a constant $L=L(m,\frA)$ such that
\[
\|F^{m-1}(E_1;t)-F^{m-1}(E_2;t)\| \le L \sum_{j=0}^{m-1}\|
 E_*^{(j)}(t)\|_2
\]
and
\[
\|\sG^{m-1}(t)\|+\|\sG^{m-1}_t(t)\|\le  L \left[ \sum_{j=1}^{m}
\| n_*^{(j)}(t)\|
+  \sum_{j=0}^{m-1}
\| E_*^{(j)}(t)\|
+ \| E_*(t)\|_2 \right],
\]
where
$\sG^{m-1}(t)= G^{m-1}(n_1,E_1;t)-G^{m-1}(n_2,E_2;t)$.
\end{lemma}
\begin{proof}
This is direct calculation based on (\ref{sm-2}) for $k\le m-1$
and on the Sobolev embeddings stated in (\ref{sobolev}).
\end{proof}
To continue with the proof of Proposition~\ref{pr:t-smo} we
introduce the variables:
\[
\widehat{E}(t)= \wE_1(t)-\wE_2(t)= E_*^{(m-1)}(t)
~~\mbox{and}~~
\widehat{n}(t)= \wn_1(t)-\wn_2(t)= n_*^{(m-1)}(t).
\]
Since $\hE$ solves the equation
\begin{equation}\label{hat-E}
  i \hE_t-(A + h^2 A^2) \hE -n_1(t)\hE +i\gamma \hE
=n_*(t)\wE_2(t)   +\sG^{m-1}(t),
\end{equation}
by Lemma~\ref{le:schr-lin} we have that
\begin{align*}
 \d \|  \hE_t\|^2 + 2\ga  \| \hE_t\|^2 & =2\Im (n_{1t}\hE +
n_{*t}\wE_2 +n_{*}\wE_{2t}  +\sG_t^{m-1}, \hE_t)  \\
& \le  \ga  \| \hE_t\|^2 +
C \left[ \| n_*^{(m)}(t)\|^2 +Q_m(n_*,E_*; t)
 \right],
\end{align*}
where
\begin{equation*}
Q_m(n_*,E_*; t) =\sum_{j=1}^{m-1}
\left[\| n_*^{(j)}(t)\|^2
+
\| E_*^{(j)}(t)\|^2\right]+ \| n_*(t)\|^2_2
+ \| E_*(t)\|^2_2.
\end{equation*}
 Thus
\begin{align}\label{hE-est}
 \d \|  \hE_t\|^2  + \ga  \| \hE_t\|^2
 \le C \left[ \| n_*^{(m)}(t)\|^2 +Q_m(n_*,E_*; t)
 \right],
\end{align}
From equation (\ref{w-nd1}) we see that $\hn$
solves equation
\[
  \hn_{tt}+ \left(A+ h^2 A^2\right)\hn+\alpha \hn_t
=F^{m-1}(E_1;t)-F^{m-1}(E_1;t).
\]
Thus
 by Lemma~\ref{le:n-lin} we obtain that the function
\[
W_0(t)=W_0(\hn_t,\hn) =
\frac12\left[ \|\hn_t\|^2+\| \hn\|_1^2+ h^2\|\hn\|_2^2\right]
+\delta \left[ (\hn,\hn_t)+
\frac{\alpha}2 \|\hn\|^2\right]
\]
with appropriate $\eta,\delta>0$ satisfies
\begin{align}\label{hn-est}
 \d &W_0(t) +\eta W_0(t)
\le - a_0  \|\hn_t\|^2+ C \|F^{m-1}(E_1;t)-F^{m-1}(E_2;t)\|^2
\nonumber
\\ & \le
-a_0 \|n_*^{(m)}(t)\|^2 +
  \sum_{j=0}^{m-2}\|
 E_*^{(j)}(t)\|^2_2 + \eps  \|\hE (t)\|^2_4+C_\eps
 \|E_*^{(m-1)}(t)\|^2
\end{align}
for any $\eps>0$.
It follows from (\ref{hat-E}) that
\begin{align*}
\|\hE\|_4^2 &\le C\left[
  \| \hE_t\|^2+ \|\hE\|^2   + \|n_*(t)\|^2_2   +\|\sG^{m-1}(t)\|^2\right],
\\ & \le  C\left[
  \| \hE_t\|^2+ Q_m(n_*,E_*; t) \right].
\end{align*}
Therefore  from (\ref{hE-est}) and (\ref{hn-est}) with appropriate
$\nu>0$ and $\mu>0$  we obtain
\begin{align*}
 \d \Big[W_0(t)  +\nu  \| \hE_t\|^2 \Big]  +\mu
\Big[W_0(t) +\nu  \| \hE_t\|^2 \Big] \le C \widetilde{Q}_m(n_*,E_*; t),
\end{align*}
where
\begin{align}\label{Q-m}
\widetilde{Q}_m & (n_*,E_*; t) = {Q}_m(n_*,E_*; t)
+  \sum_{j=0}^{m-2}
\| E_*^{(j)}(t)\|_2^2
\nonumber
\\
&\le c\left[
\sum_{j=1}^{m-1}
\| n_*^{(j)}(t)\|^2
+  \sum_{j=0}^{m-2}
\| E_*^{(j)}(t)\|_2^2+ \| n_*(t)\|^2_2
+ \| E_*^{(m-1)}(t)\|^2\right].
\end{align}
This implies the following stabilizability estimate
\begin{multline*}
\| n_*^{(m)}(t)\|^2+ \| n_*^{(m-1)}(t)\|_2^2
+ \| E_*^{(m)}(t)\|^2
\\
 \le  Ce^{-\mu  (t-s)}
\big[ \| n_*^{(m)}(s)\|^2+ \| n_*^{(m-1)}(s)\|^2
+ \| E_*^{(m)}(s)\|^2 \big]
\\  +
C\int_s^t e^{-\mu (t-\t)}  \widetilde{Q}_m(n_*,E_*; \t) d\t
\end{multline*}
for every $t>s$.
Let now
 $U_1(t)=U(t)=(n_{t}(t);n(t);E(t))$ and
 $U_2(t)= U(t+\delta)$ for some $\delta$.
In this case
\[
 n_*^{(j)}(\t)= - \int_\t^{\t+\delta}n^{(j+1)}(\si)d\si
~~~\mbox{and}~~~ E_*^{(j)}(\t)= - \int_\t^{\t+\delta}E^{(j+1)}(\si)d\si.
\]
Therefore  by the induction hypothesis we have from  (\ref{Q-m}) that
\[
 \widetilde{Q}_m(n_*,E_*; \t)\le C\delta^2, ~~~\forall\, \tau\in \R.
\]
Thus after the limit transition $s\to-\infty$ we obtain
\begin{multline*}
\| n^{(m)}(t+\delta)- n^{(m)}(t) \|^2+
\| n^{(m-1)}(t+\delta)- n^{(m-1)}(t) \|_2^2
\\
+ \| E^{(m)}(t+\delta)- E^{(m)}(t) \|^2\le
 C \delta^2,~~~\forall\, t\in \R, ~~ |\delta|<1.
\end{multline*}
In the limit $\delta\to 0$ this implies that
\[
\| n^{(m+1)}(t) \|^2+
\| n^{(m)}(t) \|_2^2
+ \| E^{(m+1)}(t) \|^2\le
 C,~~~\forall\, t\in \R.
\]
From the corresponding equations for $ n^{(m-1)}(t)$ and $E^{(m)}(t)$
we conclude that $ \| n^{(m-1)}(t) \|_4^2
+ \| E^{(m)}(t) \|_4^2\le
 C$ for all $t\in\R$. Thus the proof of Proposition~\ref{pr:t-smo}
is complete.
Hence the statement 4 in Theorem~\ref{t:attr} is proved.
\end{proof}
\subsection{Spatial regularity}\label{Sec3.5}
Now we prove the statement 5 in Theorem~\ref{t:attr}.
It follows from the structure of $F^{m-1}(E;t)$ and
 $G^{m-1}(n,E;t)$ in (\ref{F-m}) and (\ref{G-m})
and from (\ref{U-atr-bnd})
that on the attractor we have
\[
F^{m-1}(E;t)\in H^2(\Om)~~\mbox{and}~~
G^{m-1}(n,E;t) \in H^4(\Om)~~~\mbox{for every}~~ m\ge 2.
\]
Thus from (\ref{w-nd1}) and (\ref{w-nd2}) we conclude
that
\[
\Delta^2 n^{(m-1)}(t)\in H^2(\Om)~~\mbox{and}~~
\Delta^2 E^{(m-1)}(t)\in H^4(\Om).
\]
Thus by elliptic regularity (see, e.g., \cite{Lions})
we have that
\[
 n^{(m-1)}(t)\in H^6(\Om)~~\mbox{and}~~
 E^{(m-1)}(t)\in H^8(\Om)~~~\mbox{for every}~~ m\ge 2,
\]
which implies that
\[
F^{m-1}(E;t)\in H^6(\Om)~~\mbox{and}~~
G^{m-1}(n,E;t) \in H^6(\Om),~~~ m\ge 2.
\]
and so on.
This  completes the proof of Theorem~\ref{t:attr}.

\end{document}